\documentclass[10pt,a4paper,twoside]{article}
\usepackage{epsfig}
\usepackage{baltlat6}
\usepackage{array}
\usepackage{here}
\begin{document}
\ \
\vspace{0.5mm}
\setcounter{page}{1}

\titlehead{Baltic Astronomy, vol.\, , , 2014}

\titleb{STRUCTURE AND KINEMATICS OF THE POLAR RING GALAXIES: NEW OBSERVATIONS AND ESTIMATION OF THE DARK HALO SHAPE.}

\begin{authorl}
\authorb{A. Moiseev}{1,2},
\authorb{S. Khoperskov}{3,2,4},
\authorb{A. Khoperskov}{5},
\authorb{K. Smirnova}{6},
\authorb{A. Smirnova}{1},
\authorb{A. Saburova}{2} and
\authorb{V. Reshetnikov}{7}

\end{authorl}

\begin{addressl}
\addressb{1}{Special Astrophysical Observatory, Russian Academy of Sciences, Nizhniy Arkhyz, Karachai-Cherkessian Republic 357147, Russia}
\addressb{2}{Sternberg Astronomical Institute of Moscow M.V. Lomonosov State University, Universitetskii pr. 13, Moscow, 119992, Russia}
\addressb{3}{Dipartimento di Fisica, Universit\`{a} degli Studi di Milano, via Celoria 16, I-20133 Milano, Italy}
\addressb{4}{Institute of Astronomy, Russian Academy of Sciences, Pyatnitskaya str. 48, Moscow, 119017, Russia}
\addressb{5}{Volgograd State University, Universitetsky pr., 100, 400062 Volgograd, Russia}
\addressb{6}{Ural Federal University, 19 Mira street,  620002, Ekaterinburg, Russia}
\addressb{7}{St.Petersburg State University, Universitetskij pr. 28, 198504 St.Petersburg, Stary Peterhof, Russia}

\end{addressl}


\begin{summary}

The polar ring galaxies (PRGs) represent an  interesting type of peculiar systems in which the  outer matter is rotating in the plane which is roughly perpendicular to the disk of the main galaxy. Despite the long lasting study of the PRGs there is a lack of the detailed enough observational data, there are still many open questions. Among the most interesting issues there are the estimate of the flattening of the dark matter halos in these systems and the verification of the assumption that the most massive polar structures were formed by the accretion of the matter from the intergalactic filaments. The new catalog recently collected by our team using the SDSS images increased by several times the number of known PRGs.  The current paper gives the  overview of our  latest results on the study of morphological and photometric structure  of the PRGs. Using the stellar and ionized gas kinematics data based on spectral observations  with the Russian 6-m telescope we  estimate   the dark matter  halo shape in the  individual galaxies.

\end{summary}

\resthead{The polar ring galaxies structure and kinematics} {A. Moiseev et al.}
\sectionb{1}{INTRODUCTION}

This year we celebrated 110th anniversary of Boris Vorontsov-Vel'yaminov who wrote on the first Russian edition of his book  (Vorontsov-Vel'yaminov   1972): ``...it seems that  a single plane of symmetry is often absent in spiral galaxies, but   two or more   planes of symmetry co-exist which are   sometimes under the significant inclination to each  other...'' Latter Vera Rubin called such objects (including also S0 and ellipticals) as ``multi-spin galaxies'' (Rubin 1994). The polar ring galaxies  represent the one of the most famous and beautiful sort  of multi-spin objects with   an external  disks or rings rotating at the large angle relative to the central galaxy plane.

The existence of PRGs was discovered long ago.  Sandage (1961) revealed an unusual galaxy -- NGC\,2685 which seems to have two axes of symmetry. He distinguished filaments wrapping a central early type galaxy and called it ``a spindle''. In the subsequent study of this galaxy Schechter \& Gunn (1978) obtained the long-slit spectra along the major and minor axes  and concluded that this galaxy could be formed through the accretion of a galaxy or a gas cloud by the central S0 galaxy.
Interestingly J\'ozsa et al. (2009) using  the deep HI and optical observations found that NGC\,2685 is not a classical polar system since  it contains an  extremely warped coherent disk which is  inclined at  $\sim70\degr$ to the main lenticular body and coplanar with it at the external radii.

The problem of the formation of the PRGs still remains challenging. Nowadays several scenarios of the formation of the PRGs are proposed, for review and original references see Bournaud \& Combes (2003), Combes (2006, 2014). It is believed that the majority of external polar structures originate  from the direct merging of the orthogonal disk galaxies or from the accretion of the companion's matter (Reshetnikov \& Sotnikova1997). Some of the PRGs could be formed by the accretion of cold gas from the filaments of intergalactic medium (for review see Iodice 2014) .

Despite the increasing number of known PRGs there is still a lack of individual PRGs which were studied in great detail. Partly for this reason the question of the dark matter halo shape in the PRGs remains open in spite of the fact that they represent the unique objects for this study. According to the position of the PRGs on the Tully-Fisher diagram they tend to possess higher rotation velocities for given values of luminosity. It could indicate that the DM halo is flattened toward the plane of the polar component (Iodice et al. 2003; Reshetnikov 2004).  Snaith et al. (2012) performed the simulation of the formation of a PRG through the combination of  major merging and cold filamentary  accretion and  found that the shape of the DM halo of the simulated galaxy changes with radius and becomes flattened in the direction of the polar disk for the outer radii.

This paper is an expanded version of the talk presented on all-Russian conference ``Modern stellar astronomy-2014'' dedicated to the anniversaries of B. P. Gerasimovich and  B. A. Vorontsov-Vel'yaminov. Here we briefly review our recent results on the morphological and kinematics study of the PRGs including  the detailed  estimation of the dark matter (DM) halo shape.

\begin{figure} [!th]
\centerline{\psfig{figure=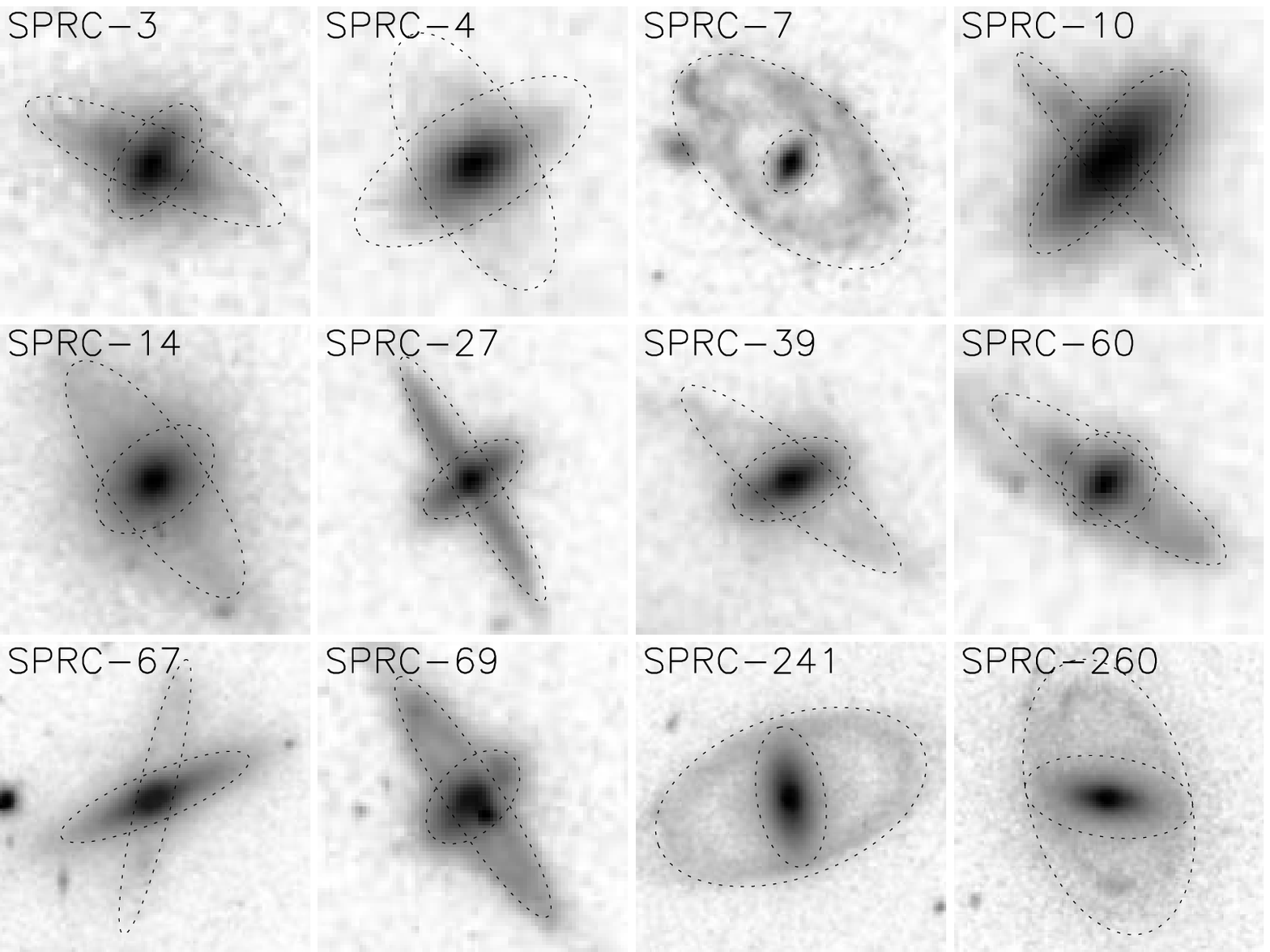,width=125mm,angle=0,clip=}}
\vspace{1mm}
\captionb{1}
{SDSS DR9 images ($g+r+i$) of the SPRC galaxies with polar structures confirmed by the kinematic measurement on the SAO RAS 6-m telescope, see references in Moiseev et al. (2014). The ellipses mark the accepted outer borders of the central galaxy disk and polar ring according to Smirnova \& Moiseev (2013).
}
\end{figure}

\sectionb{2}{NEW CATALOG OF POLAR RING GALAXIES}

The very important step in the study of the polar rings (and the collection of the first sample of PRGs) was a catalog   by  Whitmore et al. (1990) based on the galaxies photograpics  from UGC and ESO catalog. This Polar Rings Catalog (PRC) consisted of 157 objects:  for 6 galaxies the existence of the polar ring was confirmed by kinematical study,  27 galaxies were  'good candidates', while the greater part of the list was occupied by  possible candidates and related objects. Nowadays only about 20 candidates from the  Whitmore et al. (1990) catalog  have kinematic confirmation  that both the external ring and  central galaxy appear to be in a regular rotation around the same dynamical center but in the different planes.  Unfortunately
only few of them have   properties which are optimal to study the DM halo shape(`the ideal polar ring galaxy', Combes  2014): the extended and wide gas-reach polar ring/disk with the relatively small mass that the polar ring material does not perturb the shape of the galaxy DM halo.

The number of known PRGs increased significantly with release of  the  catalog  by Moiseev et al. (2011) mainly based on the results of SDSS GalZoo project. This Sloan-based Polar Ring Catalog (SPRC) contains 70
 ``the best candidates'' among which we expect to have a very high proportion of true PRGs (see the examples in Figure~1),   115 ``good  candidates'',   53 related objects (including warped discs, and mergers), and 37  possible face-on polar rings.

The SPRC provides us with an opportunity to study  the structure and properties  of PRGs using the uniform SDSS data. Finkelman et al. (2012) have shown that  the ``best'' and ``good'' SPRC galaxies preferentially inhabit low density environments compared to the other early type galaxies.  It may be explained by the destruction of the extended polar ring by close neighbours (note, that some polar rings were found also in the groups and even on the outskirt of the Virgo cluster, as NGC\,4262 considered in Sect.5). They also found that the starburst nuclei in SPRC  galaxies are significantly larger than in a control sample of early-type galaxies (ETG) and comparable with dusty ETG, while  the percentage of AGN phenomena is significantly lower than in ETG and the same with a control sample. These facts are obviously related with   specific initial condition of ``secondary event'' which forms a polar structure.

Smirnova \& Moiseev (2013) have measured  the angle between the polar ring and the central disk ($\delta$), and  the optical diameters of the both components. The most of ``the best'' SPRC rings are indeed polar, i.e. tend to lie in the plane that is almost perpendicular to the central body -- $\delta>70-80\degr$. However moderately inclined outer structures ($\delta=40-55\degr$) are observed   in about 6\% of objects which probably indicates their short lifetime.  It's interesting to compare the relative size of the outer polar rings with the circumnuclear polar structures from Moiseev (2012) list compiled from literature.   The SPRC objects and inner polar structures  form a single family in the distribution of diameters normalized
to the optical size of the galaxy. At the same time, this distribution is bimodal, as the number of objects with
$d_{ring} \approx(0.4-0,7)d_{disk}$ is small that is most likely due to the fact that the stability of polar orbits in the inner regions of galaxies is supported  by the bulge or the bar, while in the outer radii polar rings are maintained by the gravity of the DM halo.

\sectionb{3}{SPRC GALAXIES:THE PHOTOMETRIC PROPERTIES}

Reshetnikov \& Combes (MNRAS, submitted) present the results of photometric 2D
decomposition of the SDSS images of  46   the ``best'' SPRC candidates  and 4 kinematically-confirmed PRGs from  PRC   using the GALFIT code  (Peng et al. 2010). We fit a two-component model (host galaxy + ring) to all galaxies in the sample. Both components were described  by a single
S\'ersic function with seven free parameters (object centre, total magnitude,
effective radius $r_e$, S\'ersic index $n$, ellipticity, position angle).
In the  cases of  non edge-on ring we approximated the ring by the inner-truncated model in order to describe  clearly the central  hole in the ring component.

The  main results are as follows:

(i) The central (host) galaxies of the PRGs are non-dwarf  sub-$L^{\ast}$ galaxies  with average absolute magnitude $< M_r > = -20.3 \pm 1.0$.

(ii) The distribution of S\'ersic indices of the host galaxies  indicates a broad range
of morphologies, from disk-dominated ($n < 1.5$) to bulge-dominated ($n > 3$).
The relative
fractions of disk-dominated and bulge-dominated brightness distributions are 16\%, 28\%,
respectively, in the $r$ band. These fractions are consistent with previous findings
that late-type galaxies are less frequent among PRGs in comparison with early-type.
 It is important to note that $n$ is not a direct measure of the galaxy type and
it does not translate one to one to the bulge-to-total ratio. But the single S\'ersic
index is a reasonable statistical characteristic to separate late-type and early-type
galaxies (Bruce et al. 2012).

(iii) The mean colors of the host galaxies in the sample are  typical
for early-type galaxies: $<g-r> = +0.74 \pm 0.16$, $<r-i> = +0.41 \pm 0.09$.

(iv) The distributions of the host galaxies characteristics on the planes $M_r$--$r_{eff}$
and ``effective surface brightness -- $r_{eff}$'' (Kormendy relation) are shifted by $\sim 1^m$
to fainter magnitudes in comparison with typical E/S0 galaxies.
 This shift shows that the structure of the PRGs hosts can be different from the
structure of ordinary early-type galaxies. It means that the use of standard scaling
relations (Fundamental Plane, Faber--Jackson relation, etc.) for the PRGs hosts
should be made with caution.

(v) Polar structures are, on average,  fainter ($< M_r > = -18.90 \pm 1.28$) and bluer
($<g-r> = +0.61 \pm 0.25$, $<r-i> = +0.33 \pm 0.22$) than their host galaxies. Nevertheless
 in the most galaxies, stellar mass of polar component is not negligible
in comparison with stellar mass of host galaxy.

\begin{figure} [!th]
\centerline{
\psfig{figure=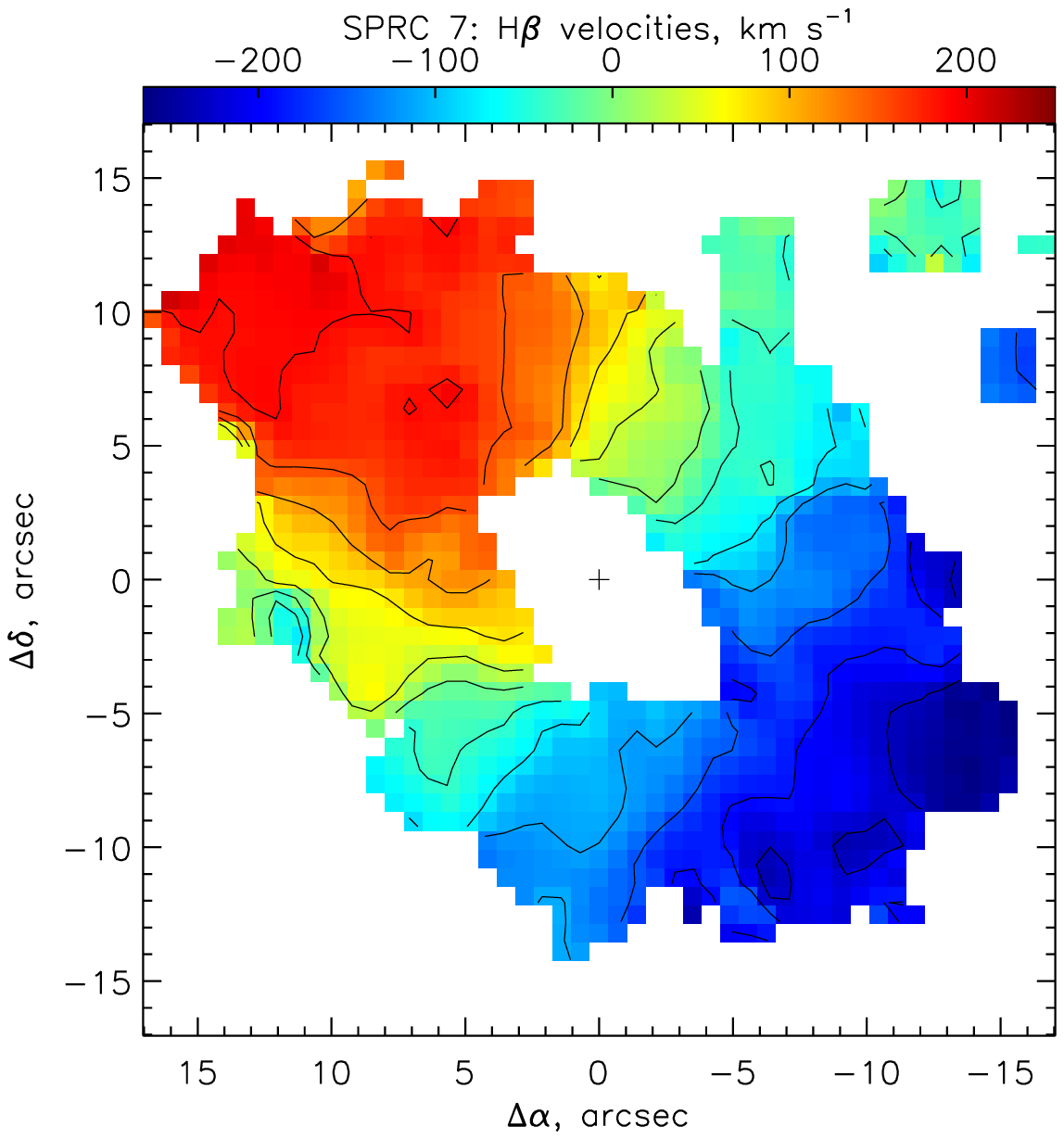,height=45mm,angle=0,clip=}
\psfig{figure=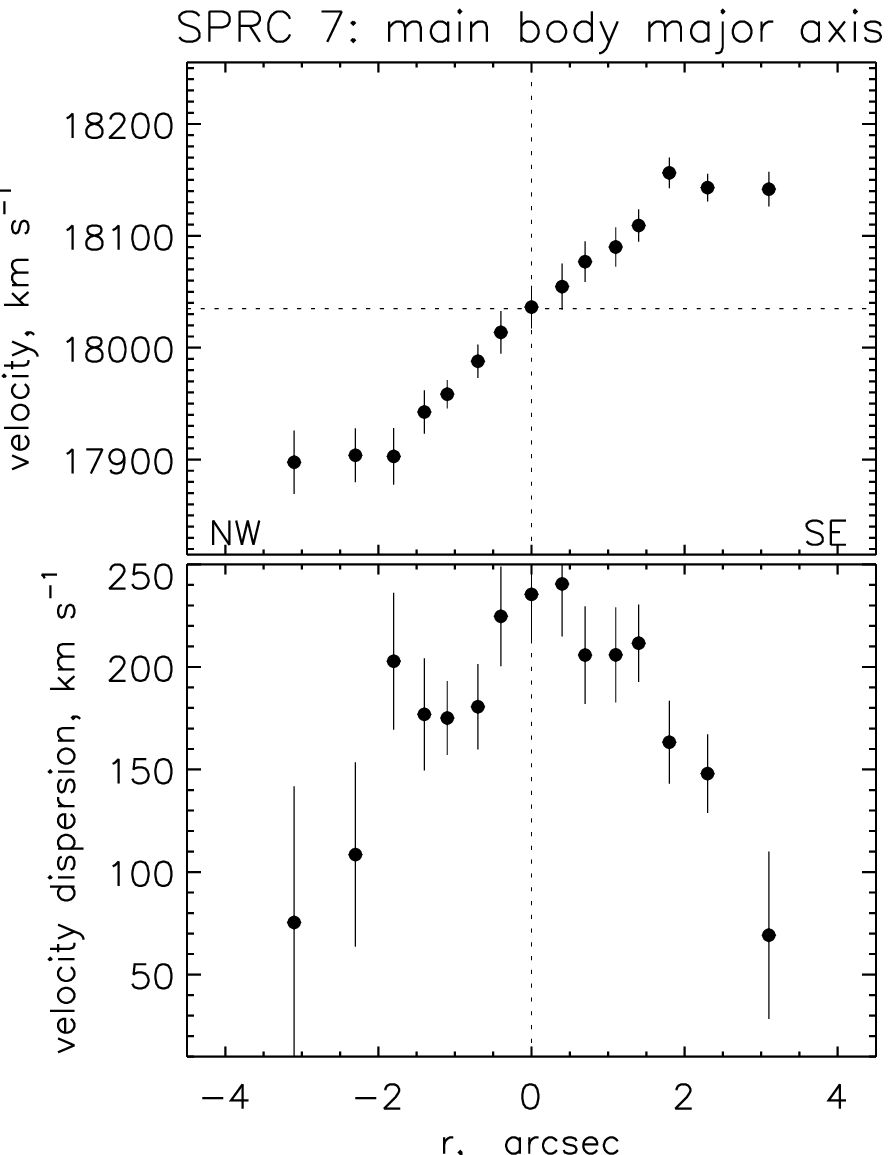,height=45mm,angle=0,clip=}
\psfig{figure=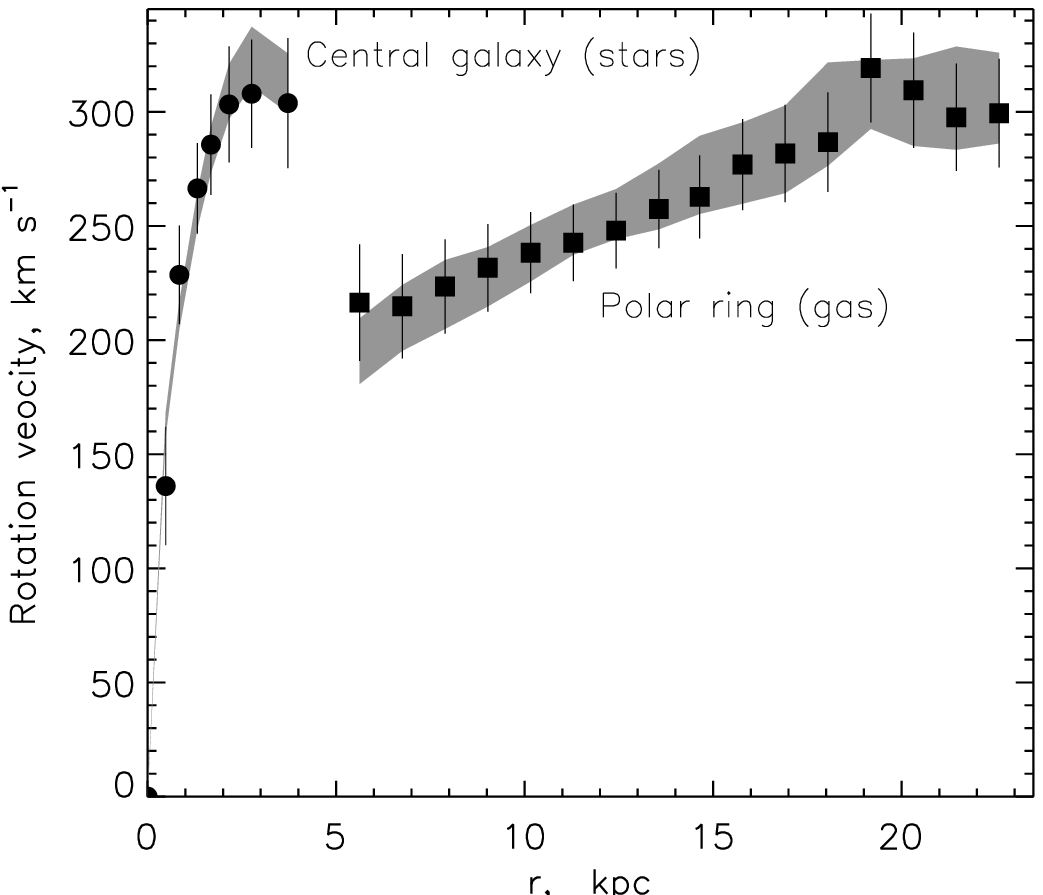,height=45mm,angle=0,clip=}
}
\vspace{1mm}
\captionb{2}
{The kinematics of the SPRC-7 galaxy studied at the SAO RAS 6-m telescope: the  ionized gas velocity field of the polar disk (left), the line-of-sight velocity and velocity dispersion distributions for the stars along host galaxy major axis obtained from long-slit spectroscopic observations (middle),  the fit of the  rotation curves of the both components  with the model  including NFW halo with $c/a=1.5$ ( right) } \end{figure}

\sectionb{4}{KINEMATICS OF  GAS AND STARS OF THE PRGS}

We performed the spectroscopic observations of SPRC galaxies  at   the prime focus  of the Special Astrophysical Observatory of the Russian Academy of Science  6-m telescope    with the  multi-mode focal reducer SCORPIO (Afanasiev \& Moiseev, 2005)  and its new version SCORPIO-2  (Afanasiev \& Moiseev, 2011) in the scanning Fabry-Perot interferometer and long-slit spectroscopic modes.  These    observations of SPRC objects first of all are aimed to  confirm that the external  component  in SDSS images is not a trivial projection of a distant galaxy, but  represents a  matter  rotating on polar orbits. Now  15 candidates including the literature data cited in Moiseev et al. (2011) already have  received the kinematic confirmation, 12  of them  were confirmed by  observation at   the 6-m telescope   (Figure~1). Other goals of our observations are following:

(i) Estimation of the ionized gas metallicity in polar component that is one of the important parameter for  the verification of cold accretion scenario for formation of wide polar disks (Spavone et al. 2010).

(ii) Measurement of the stellar  velocity dispersion of the central galaxy in order to compare their  Faber-Jackson relation with other early type galaxies, for motivation see Iodice (2014).

(iii) Study of the shape of the DM halo  using rotation curves observed in two orthogonal planes.

Recently  Combes et al. (2013)  presented the molecular gas rotation amplitude  for the small sample of SPRC galaxies observed with IRAM 30 m telescope.  They confirmed the offset of PRGs in the velocity-magnitude diagram  from the Tully-Fisher relation of normal spirals, which is usually interpreted as a rotation on the elliptical orbits in DM halo flattened to the polar plane (Sect.~1). However   the usage   of the classical scaling relation for peculiar objects is  controversial, therefore direct and independent  estimation of DM shape is still need. We have performed this sort of analysis for three SPRC galaxies, the results are presented below.

\begin{figure} [h]
\centerline{
\psfig{figure=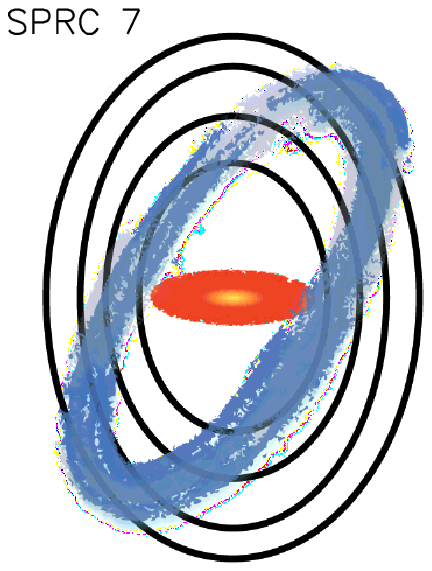,height=45mm,angle=0,clip=}
\psfig{figure=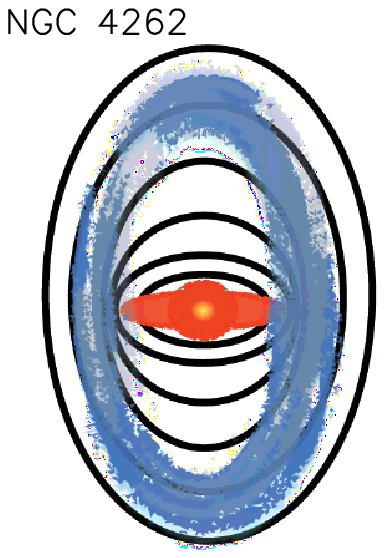,height=45mm,angle=0,clip=}
\psfig{figure=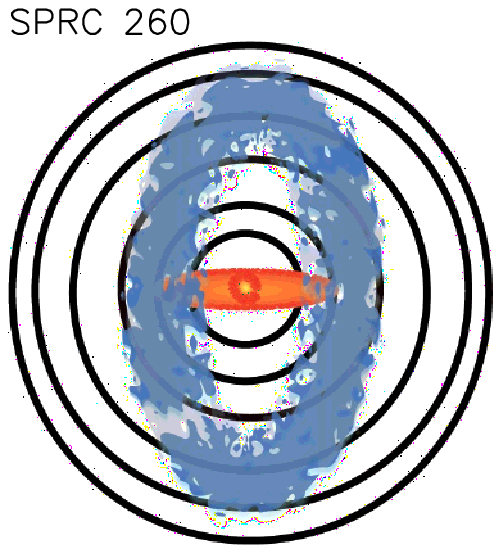,height=45mm,angle=0,clip=}
}
\vspace{1mm}
\captionb{3}
{The distribution of the gravitational potential of the dark halo (contours) according to the numerical calculations in the  galaxies considered. The central disk and the polar ring are shown schematically. }
\end{figure}

\sectionb{5}{THE ESIMATES OF THE DM SHAPE OF THE PRGS}

\label{sect5}

Figure~2 shows the example of observational data used in our analysis of SPRC-7 kinematics. The polar ring ionized gas rotation curve was derived from the titlted-rings analysis of the velocity field created from  Fabry-Perot interferometer observations (Brosch et al. 2010). The circular rotation curve for the central galaxy was evaluated from the rotation curve of the stellar component  along the galaxy major axis  corrected for the asymmetric drift using radial distribution of the line-of-sight velocity dispersion. The similar data set was collected for SPRC-260 galaxy, while for NGC\,4262 (SPRC-33) the   HI data published by   Oosterloo et al. (2010) were used to analyze the polar component rotation and gas density distribution.  Apart from the data on the motion of gas and stars, the detailed model of galaxies has also included the information about the radial profiles of the surface  density  of the stellar component according to the SDSS images.

We constructed the models that reproduced the observed kinematics of both the main body and the polar component of the considered PRGs; for more details see Khoperskov et al. (2013, 2014).  The shape of the dark halo in the model was chosen so as to achieve the best fit to the observed rotation of both galaxies themselves and their polar rings. According to our study SPRC-260 possesses a triaxial DM halo with the scale lengths ratios:  $c/b=0.95$, $a/b=1.1$.

It's interesting to compare SPRC-7 and NGC\,4262. The first system   is one of the largest (the ring diameter is about 50 kpc) and the most distant among the confirmed PRGs, a massive ring here consists of gas and stars and rotates not perpendicularly to the plane of the galactic disk, but rather at an angle of $73\degr$ degrees to it. NGC\,4262 belongs to the  Virgo cluster, the ring of about 30 kpc in diameter contains almost no stars, it is mainly composed of the HI gas and is oriented almost exactly orthogonally to the central galaxy. The studied galaxies notably differ from each other. Not surprisingly, the dark haloes in them are dramatically different too. In SPRC-7 the halo is noticeably flattened  towards the polar ring plane (Figure~3). For this galaxy we used two different DM density profiles -- NFW (Navarro et al. 1996) and isothermal (Burkert 1995). For the NFW DM halo we obtained the axis ratio $c/a=1.5\pm0.2$ while for the isothermal distribution -- $c/a=1.7\pm0.2$.

For NGC\,4262 the situation is more complicated. The models with the DM halo axis ratio which is constant with radius failed to reproduce the observed kinematics of gas and stars. Thus we used the model in which the axis ratio $c/a$ could vary with radius amounting to about 0.4 in the inner regions and 1.5-2.3 in the outer regions (see Figure~3). This is the first galaxy (except for Milky Way), where the variation of the shape of the dark halo with radius is reliably defined, as it was predicted by some theoretical models of galaxy formation (Snaith et al. 2012).

\sectionb{6}{CONCLUSION}
The capabilities of the project Galaxy Zoo and SDSS data enabled to increase the number of good candidates to PRGs by a factor of three. The number of PRGs with kinematic confirmation also rises.

We showed on the examples of individual PRGs that one can estimate the shape of the DM halo in the polar ring galaxies. However to do that  one needs accurate enough observational data.

According to our estimates of the DM halo shape in the PRGs the flattening of the halo could be different in different objects. It could indicate the absence of one universal formation scenario for all PRGs.

\thanks{ }
The work was supported by the RFBR grant  13-02-00416  and the "Active Processes in Galactic and Extragalactic Objects" basic research program of the Department Physical Sciences of the RAS OFN-17 and also it was partly supported by St. Petersburg State University research 
grants 6.0.160.2010, 6.0.163.2010, and 6.38.71.2012.  A.M. and A.S. are grateful to the nonprofit 'Dynasty' Foundation.  The observations at the 6-m telescope were carried out with the financial support of the Ministry of Education and Science of Russian Federation (contracts no. 16.518.11.7073 and 14.518.11.7070).

\References

\refb   Afanasiev V.~L., Moiseev A.~V. 2005, Astronomy Letters, 31, 194

\refb   Afanasiev V.~L., Moiseev A.~V. 2011, Baltic Astronomy, 20, 363

\refb   Bournaud F., Combes F. 2003, A\&A, 401, 817

\refb   Brosch N., Kniazev A.~Y., Moiseev A., Pustilnik S.~A. 2010, MNRAS,
  401, 2067

\refb  Bruce V.~A., Dunlop J.~S., Cirasuolo M., McLure R.~J.  et al. 2012, MNRAS,
  427, 1666

\refb  Burkert A. 1995, ApJ, 447, L25

\refb   Combes F. 2006, in {\it Mass Profiles and Shapes of Cosmological Structures}, eds. G. Mamon et al., EAS Publications Series, 20, 97

\refb  Combes F. 2014, in {\it Multi-Spin Galaxie}, eds. E. Iodice \& E. M. Corsini, ASP Conf. Series, 486,  207

\refb   Combes F., Moiseev A., Reshetnikov V. 2013, A\&A, 554, A11

\refb   Finkelman I., Funes J.~G., Brosch N. 2012, MNRAS, 422, 2386

\refb  Iodice E. 2014, in {\it Multi-Spin Galaxie}, eds. E. Iodice \& E. M. Corsini, ASP Conf. Series, 486, 39

\refb Iodice E., Arnaboldi M., Bournaud F., Combes F. et al. 2003, ApJ, 585, 730

\refb  J\'ozsa G.~I.~G., Oosterloo T.~A., Morganti R., Klein U., Erben T.
  2009, A\&A, 494, 489

\refb Khoperskov S., Moiseev A., Khoperskov A. 2013, Memorie della Societa
  Astronomica Italiana Supplementi, 25, 51

\refb  Khoperskov S.~A., Moiseev A.~V., Khoperskov A.~V., Saburova A.~S. 2014,
  MNRAS, 441, 2650

\refb  Moiseev A.~V. 2012, Astrophysical Bulletin, 67, 147

\refb  Moiseev A.~V., Smirnova K.~I., Smirnova A.~A., Reshetnikov V.~P. 2011,
 MNRAS, 418, 244

\refb  Moiseev A., Egorov O., Smirnova K. 2014, in {\it Multi-Spin Galaxie}, eds. E. Iodice \& E. M. Corsini, ASP Conf. Series, 486, 71

\refb  Navarro J.~F., Frenk C.~S., White S.~D.~M. 1996, ApJ, 462, 563

\refb  Oosterloo T., Morganti R., Crocker A., J\"utte E., Cappellari M. et al. 2010, MNRAS, 409, 500

\refb  Peng C.~Y., Ho L.~C., Impey C.~D., Rix H.-W. 2010, AJ, 139, 2097

\refb  Reshetnikov V., Sotnikova N. 1997, A\&A, 325, 933

\refb  Reshetnikov V.~P. 2004, A\&A, 416, 889

\refb Rubin V.~C. 1994, AJ, 108, 456

\refb  Sandage A. 1961, {\it The Hubble atlas of galaxies}, Washington: Carnegie Institution

\refb  Schechter P.~L., Gunn J.~E. 1978, AJ, 83, 1360

\refb Smirnova K.~I., Moiseev A.~V. 2013, Astrophysical Bulletin, 68, 371

\refb  Snaith O.~N., Gibson B.~K., Brook C.~B., Knebe A. et al. 2012, MNRAS, 425, 1967

\refb  Spavone M., Iodice E., Arnaboldi M., Gerhard O. et al. 2010, ApJ, 714, 1081

\refb  Vorontsov-Vel'yaminov B.~A. 1972, in {\it Extragalactic astronomy}, textbook
  for students at universities, Moscow, USSR

\refb Whitmore B.~C., Lucas R.~A., McElroy D.~B., Steiman-Cameron T.~Y. et al. 1990, AJ, 100, 1489

\end{document}